# Towards an Understanding of Microservices

Dharmendra Shadija and Mo Rezai
Department of Computing
Sheffield Hallam University, UK
{d.shadija},{m.j.rezai}@shu.ac.uk

Richard Hill
School of Computing and Engineering
University of Huddersfield, UK
r.hill@hud.ac.uk

*Abstract*—**Microservices architectures are a departure from traditional Service Oriented Architecture (SOA). Influenced by Domain Driven Design (DDD), microservices architectures aim to help business analysts and enterprise architects develop scalable applications that embody flexibility for new functionalities as businesses develop, such as scenarios in the Internet of Things (IoT) domain. This article compares microservices architecture with SOA and identifies key characteristics that will assist application designers to select the most appropriate approach.**

## I. INTRODUCTION

Software engineering as a discipline is mature [1] and many approaches to developing software architectures have been proposed. Our focus is upon service-based approaches, in particular microservices. Service-Oriented Architecture (SOA) has been established for some time now and we have chosen to undertake a comparative study between SOA and microservices, to enable greater understanding of the relative characteristics of each approach.

### A. The context for services

A clear methodology for structuring application logic first appeared with the advent of Jackson Structured Programming (JSP) [2]. JSP encouraged the maintenance of a library of subroutines, each of which would do one thing well (cohesion), for example printing a text string to the screen. In doing so, JSP promotes modularity and reuse of code blocks. Pervasive adoption of Object Orientation (OO) was the next paradigm shift [3]. OO focuses on creating code block abstractions (called objects) as a set of services that could be called by clients (other objects) [4]. This abstraction enables the object internals to be complex, often using services of other objects for business logic, data calculations and data transformation. Objects also encapsulate data and control access to data, for reading or for changing the encapsulated data [4]. An example of an object would be an `IncomeTaxCalc` object in an HR system. This object would provide a service for calculating income tax on an employee's salary. It could encapsulate details of the personal taxation data provided by the government of the employee's country of residence. It may not provide access to the encapsulated data as no other object in the system is likely to require access to this data. This object's service comprises of business logic and data calculation. A client of this object may be the `WageSlip` object, to create the data necessary for an individual report of an employee's contributions towards taxation. Another feature of OO that is pertinent to this discussion is messaging between the objects [4]. Objects publish well-specified interfaces and it is through these interfaces that services of objects are called by the clients. An interface to an object identifies the object, the block of code within the object that is to execute, parameters for the code, and any return type [4]. Abstraction can be described as the separation of interface of the object from the internal implementation of the object. By abstracting how an object is used from how an object works, dependency of the client upon the object is reduced [4]. Modularity is another inherent characteristic of object oriented systems. The aim of modularity is to decompose the application into highly modular objects, each of which encapsulates coherent functionalities that enable maximum reuse of the objects. Code reuse is ultimately the payback in OO. Good modularity and well-designed encapsulation results in highly reusable objects. As a consequence, object based development is generally considered to be system development at a low level of granularity [4]. Component-based system development was the next significant development. In an attempt to work at higher levels of granularity, a number of objects are packaged together as a component. The idea is that working with components (as opposed to objects) would increase productivity [4] as it would be working at relatively higher level of granularity, and therefore be easier to translate form and to business logic. `WageSlip` would be an example of a component that would encapsulate a number of objects such as `IncomeTaxCalc`, `NationalInsuranceCalc` and `PensionCalc`. Reuse of `WageSlip` would result in higher productivity payback.

### B. Service Oriented Architecture

In Service Oriented Architecture (SOA), a service encapsulates a number of components into a single interface to provide a discrete business function. For example, a service to check a share price on the financial exchange would support all functions to do with managing the checking of the share dealing account. In this respect services work at a higher level of granularity than the components [5]. A component could be a service if it is wrapped in a service layer. For example a `WageSlip` component could be exposed as a service. The aggregation components build an application and this is not the case with services. A service is consumed through late binding at runtime [34]. In the case of distributed systems, protocols that are used to access components cannot easily pass through enterprise firewalls. Service layers enable industry standard, widely accepted protocols, that simplify access to the service thus promoting interoperability.

## C. Multi-tier architectures

Fig 1 illustrates a traditional multi-tier architecture. The core components are as follows:

- Data tier. Different data sources such as relational databases, XML databases, MS Excel spreadsheets, object databases, etc.

- Integration tier. Managing the connections to data sources (connection classes), and the execution of queries against each connection. This tier is decoupled from the rest of the tiers. There should be no business logic in this layer. Program code must be to do with data access only.

- Business tier. Classes in this layer carry out business logic. Business logic could be bespoke for the application, for instance special requirements or dealing with legacy applications. Business logic could be standard libraries from a third party if available. There is no database utility in this layer.

- Presentation tier. This layer encapsulates all of the presentation applications. Fig 1 depicts an MVC website, but the presentation tier can be extended to include web forms, mobile applications, desktop applications, etc.

- Web services tier. If required, business layer classes can be exposed as web services. In that respect a web service becomes another type of presentation tier but without a user interface and merely a data service and/or business processing service.

Such an architecture promotes the reuse of program code (via objects), although the emergence of more flexible business models is now creating a demand for architectures that can scale more freely than the traditional multi-tier model.

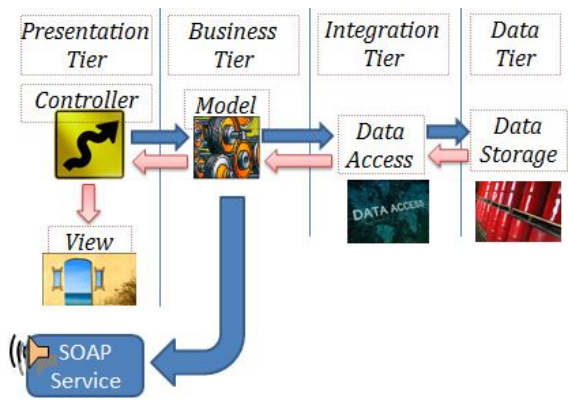

Figure 1. Illustration of a SOAP Service in relation to a Model View Controller (MVC) application architecture.

## II. DEFINING MICROSERVICES

At present, there is considerable interest in microservices architectures, amongst academic and industrial communities alike. Many explanations of microservices have emerged in recent years as communities attempt to propose a definition. As we work towards a definition, many inconsistencies and misnomers become apparent.

One such example is that of the meaning of 'micro'; some literature reports that a microservice is very small, and therefore such an architecture makes use of many, small services. As we have discovered, 'micro' is a relatively ambiguous term, that does not always describe the size of a service, as these can vary. Therefore, we have examined a range of literature that is relevant to microservices architecture, in order to assist the development community who might be considering using microservices architectures for their applications.

Dragoni et al [6] [7] define microservices as:

"A microservice is a cohesive, independent process interacting via messages."

This definition describes two key features of microservices. First, microservices should be highly cohesive units; they should do one thing well. Second, microservices must be able to execute their own processes which allows for independent deployment. However, the same definition might also be applied to an OO class.

Dragoni et al [6], [7] also offers a definition for a microservice architecture:

"A microservice architecture is a distributed application where all its modules are microservices."

Microservice architecture is about a service addressing a single business capability, with a clearly defined interface. In a microservice architecture a series of microservices are chained together to perform a bigger business function. To enable these characteristics, each microservice has its own data model and a class model. Fig 3 below shows a sample application which utilises a microservices architecture. Adrian Cockcroft defines a microservice as:

"loosely coupled service in a bounded context."

This definition refers to 'bounded context' which is derived from the Domain Driven Design [8] literature. A bounded context captures the key properties of a microservice architecture: the focus upon business capabilities, rather than program code decomposition and reuse. Such a perspective supports the capture and modelling of requirements in complex multi-agency domains such as the delivery of community healthcare [9], [10] or applications for the Internet of Things [11]. Related business functionalities are combined into a single business capability which is then implemented as a service. However, this definition does not provide any insight into the level of granularity required for the functionality to be branded as a microservice. For example, at what point does service decomposition become a method call upon another object? This is a pertinent question for designers of applications that exploit Internet of Things infrastructure [12], [13]. Another definition of microservice from Sam Newman states [14] that microservices are:

"Small autonomous services that work together, modelled around a business domain"

This suggests the importance of service autonomy, underlining the need for a service to own its own data model. Daniel Bryant [15] states that a microservice should be "... designed around the single responsibility principal." We infer from this the reference towards responsibility from the perspective of business requirements. Johannes Thones defines a microservice as [16]:

"a small application that can be deployed independently, scaled independently, and tested independently and that has a single responsibility."

This definition addresses a number of characteristics of a microservice, namely that the microservice should be self-reliant, flexible and fault-tolerant. In addition to the above this also highlights that microservices should have a single responsibility. Again, like many definitions that are being proposed, the level of granularity is still an area for further exploration. Some of the key architectural concerns that microservice architecture aims to address are around scalability, being able to deploy updates to microservices independently, and lightweight mechanisms around orchestration and choreography. Microservices communicate using Representational State Transfer (REST) or Message Queue (MQ). This light weight communication mechanism suggests that a microservice architecture is more tolerant of physical infrastructure that has distributed computation and storage. The continued growth of wireless devices places considerable demands upon architectures that cannot scale sufficiently to flexibly adopt new resources as they become available. It follows that a microservice architecture can bring its own challenges. Since each service has its own data model, replication of that data is necessary across a number of data stores; "Replicating data in real time is a difficult problem for which no good, general approach currently exists". [17]

in terms of being able to manage the whole architecture. There are considerable benefits to be had with coarser-grained services, not least for the translation of business requirements into application design. Therefore, the level of granularity of a service is an important part of microservice architecture. From Newman [14]:

"Avoid approaches like enterprise service bus or orchestration systems, which can lead to centralization of business logic and dumb services. Instead, prefer choreography over orchestration and dumb middleware, with smart endpoints to ensure that you keep associated logic and data within service boundaries, helping keep things cohesive."

A key difference between XML Web Service based architecture and microservice architecture is the decreased reliance on heavyweight middleware. In an XML Web Services based architecture, web services are glued together using robust mechanisms such as Enterprise Service Busses (ESB). ESB advocates cite the centralisation of application integration as being a significant advantage, where smaller services provide functionality for the ESB to deal with how they are integrated, routed, authenticated and ultimately, deciding how to apply business rules. However, the robust nature of an ESB component in an architecture serves also to constrain any application flexibility in the future. This, in turn, impacts upon how an application can deal with changing business needs. In contrast, a microservice architecture specifies end points with the associated business logic. As the application grows, microservices can be updated in an isolated manner and deployed without affecting the rest of the application. This is not always the case with OO architectures that promote object reuse. By decentralising the application architecture, software developers can write more logic within the microservices, and communication between services is kept simple using REST calls.

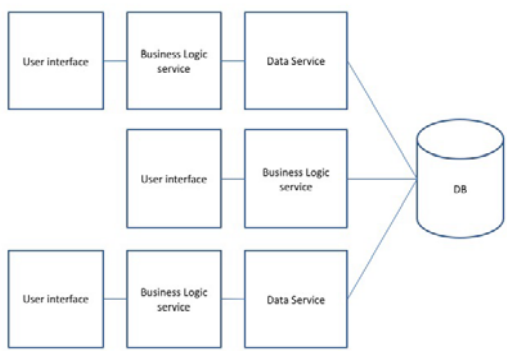

Figure 2. Example of application architecture using Service Oriented principles.

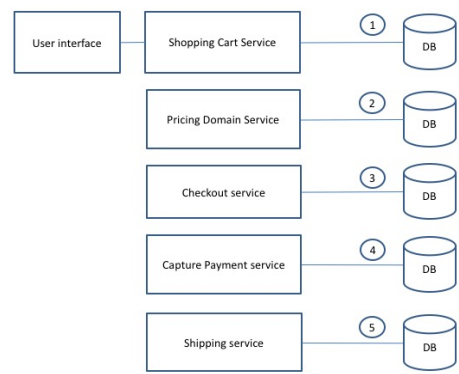

Figure 3. Example of application architecture using microservices principles.

One of the misconceptions associated with microservices is that they should be small. Microservices could be as small as a method implementation, however by having such fine grained microservices it introduces issues

As an ESB dealt with lot of the logic around service routing, data and protocol transformation [18] web services could be quite small (fine grained) as per Fig 2. For example, updating a customer record in an e-commerce application. However, with microservices, end points have to be smart [19] and services are coarser grained. Fig 3 illustrates the potential microservices in an e-commerce application: shopping cart service, checkout service, pricing service, payment processing service, fraud detection service and order fulfilment service. As the services are coarse grained it facilitates loose coupling between the services. Eric Evans advocates that a microservice approach could be looked at "from a software design perspective" [20].

Microservices architecture started with the goal to be able to deploy smaller parts of software independently without affecting rest of the application [20]. However, this has evolved and started to influence the way software is architected from the outset. Microservices therefore suit evolutionary design, where the business anticipates that certain functions may fail in the future. Business models that are scalable need applications that can be reconfigured and augmented as scenarios evolve. Since each microservice is a small business process, and because it represents a small aspect of business functionality, it is easy to replace or change the work flow. A web service based approach is more challenging in this respect, as the focus on object reuse means that changes can often affect many disparate parts of the application.

## III. CHARACTERISTICS OF A MICROSERVICES ARCHITECTURE

Lewis et al advocate a number of characteristics for a microservice architecture [19]. This section evaluates the characteristics in comparison to SOAP service architectures and is summarised in Table 1.

### A. Modularity of services

The objective of modularity in software development is to decompose the system into more manageable components. Effective decomposition renders components that are easier to deploy, to replace, and to change. SOAP services encourage loose coupling through the provision of interfaces that other applications use to consume a service. Similarly, systems are 'componentised' into a number of microservices. Microservices expose interfaces which ensure loose coupling between the microservices [6]. Componentisation in microservices also provides us with the ability to make changes to a component and only redeploy the changed component as opposed to the whole application. Microservices are therefore a move away from multi-tier architectures towards more flexible architectures. Microservices encapsulate all the resources that they require to function. One of the aims of the architecture is to facilitate scalability through virtualising a resource (the microservice) [7]. Effective modularity of the service therefore becomes realisable when it comes to scalability.

### B. Organised around business capability

This is not only about what the services do and how they are architected, but also about the constitution of the teams that build the services. In the case of typical multi-tier applications, code and usually teams are organised around functional areas. Changes to system requirements will have a consequence for cross-team communication requirements and work allocation. There is also the risk of embedding logic and/or data in layers that teams have access to. SOAP Services are constructed as an additional communication layer on existing logic, typically of a multi-tier architecture. Consequently, these services inherently suffer the detriments of injurious changes to the underlying system.

This is not the case with microservices. Microservices are not an additional communication layer. It is an application architectural style. The system is decomposed as a number of microservices, each one organised around a business capability [7]. An example of a business capability may be 'Shopping Cart Management'. This microservice would encapsulate all the related and appropriate functionalities such as 'Add to Cart', 'Remove from Cart', 'Go to Cart', 'Persist Cart', 'Retrieve Cart', etc. For better scalability, microservices also encapsulate all of the required resources including business logic and data services.

### C. Products not projects

With microservices, the design focus shifts towards business capability or the product. A business domain system is decomposed as a series of business sub-domain systems [21], each of which are further decomposed into microservices. This is in contrast with more traditional system development process where a number of projects are established to address different parts of the system. For example, we may have UI team working on a UI project and Database team working on Data Project and system, in its entirety, is designed, built, tested and placed in production. Microservice architecture advocates decentralised processes. Each process is for developing part of the system, which may be one microservice. One benefit of this approach is the loose coupling between microservices. In turn, loose coupling between the microservices expedites an evolutionary system development process and enhances future extensibility of the system.

### D. Smart endpoints and dumb pipes

Microservices are smart endpoints because they encapsulate all the resources that they require for them to function effectively. Pipes or communication between the end points is through messaging. One other tenet of Service Oriented Architecture is the decoupling of what the service does from how to communicate with the service. This is referred to as the separation of 'what' from the 'how'. The underlying idea is that communication with a service necessitates extensive use of changeable technology, and decoupling the messaging mechanism from the service will result in improved longevity of the service. In light of this, SOAPServices manifest this in the schema for their Web Services Description Language (WSDL) description. SOAP is used as the messaging mechanism between the SOAPService and its client application. SOAP is a lightweight messaging protocol that is further decoupled from the mechanism for transport albeit it usually makes use of the ubiquitous HTTP. The fundamental difference between Microservices and SOAPServices is in the use of middleware for business process. SOAPServices extensively use heavyweight middleware for orchestrating services. One example of such middleware is Enterprise Service Bus (ESB).

TABLE I. COMPARING KEY CHARACTERISTICS OF XML WEB SERVICES AND MICROSERVICES.

| Characteristic | XML Web Services | Microservices |
|---|---|---|
| Design motivations | Interoperability between heterogeneous systems [25] | Scalability of evolutionary applications |
| Messaging protocols supported | SOAP and REST | REST and MQ such as Rabbit MQ [29] |
| Message payload | No limit on the message payload | No limit on the message payload |
| Orchestration | Heavy weight using ESB or WS-* standard (WSBPEL) [26] [27] | Kubernetes [30], Docker's built-in Swarm Mode [31] Orchestration not preferred approach in microservices [14] |
| Primary use | Inter and intra-organisation communication | Primarily for intra-organisation communication |
| Choreography | Using WS-CDL | Choreography preferred approach in microservices [14] |
| Granularity | Fine grained [28] with ESB | Coarse grained as messaging between applications is lightweight |

In contrast, microservices are choreographed using RESTful protocols. Because of this SOAPServices are sometimes referred to as 'big' services and the use of the work 'micro' may be a reaction to this. Dumbness in microservices is a reference to the use of lightweight message bus [22] such as ZeroMQ [23] for simple reliable asynchronous messaging between microservices.

*E. Decentralised data management and governance*

In a microservice architecture, data is decentralised and distributed between the constituent microservices. This is in contrast with traditional application architecture and centralisation of data, usually in a relational database. This causes a number of issues. Each microservice is a solution to a business capability in a sub-domain and works with a conceptual data model for that sub-domain.

Moreover, with decentralisation of governance, the design and development of microservice is devolved to a team. A consequence of decentralisation would be lack of a unified data model for the system. As well as the different data models, Microservices may opt for different data storage systems including relational database systems, file systems, etc. Decentralising data decisions and data management has its implications. Distributed databases make use of transactions for managing updates.

IV. CONSTRAINTS AND LIMITATIONS

Alongside the potential of microservices architecture, there are a number of constraints that are imposed by this approach. First, the focus upon domain understanding means that the enterprise/software architect must be able to specify the appropriate bounded contexts for a service.

Any misunderstanding at this stage will result in services being built that might not be sufficiently cohesive, or the messaging between services might be too abundant. This will build-in to the design additional costs for the future.

Second, the principle of resilience for each microservice places additional resource demands upon the notion of scalability; conceptually a microservices architecture is much more scalable than one based on SOAPServices, however the overhead of monitoring each service creates a demand for more processing cycles and data storage than would be required with an equivalent SOA approach. As businesses take advantage of utility computing and transfer their intelligence processes to clouds [24], the additional overhead maybe absorbed by elastic compute resources, but this is still an issue that has to be considered. Third, the SOA approach is still 'purer' in that contracts between objects can be completely de-coupled (albeit at a finer grain) and the interoperability between web services is much easier to facilitate.

V. CONCLUSIONS

Decomposition of a product into microservices requires the designer to contemplate a manageable size for the microservice. Use of the word 'micro' in this context is often taken too literally and can be misleading. In addition, the term 'service' tends to encourage a direct comparison with SOAPServices.

However, a SOAPService is a communication layer on top of the business logic of an application. A microservice is a Service Oriented Architectural style for the application. Microservices are built around business capability. One tenet of SOA is that a service must be of a tangible benefit to the consumer. SOA does not assert a size for the service but tangibility in this context could mean 'usefulness', suggesting that if a service is not useful, consumers will not demand it.

Tangibility to the consumer, more often than not, hints at larger and more substantial services with the potential to remove a sizeable burden. For example, a payment micro service should oversee all payment utilities including take payment, make refund, deal with payment enquiry, etc.

Hence, microservices are generally coarse-grained. SOAPService is an integration technology whereas microservices is the application architecture. They do have some similarity as they are both subsets of SOA. In enterprise application development, there is scope for both approaches to be utilised, particularly when there is a foreseen business need for an application to scale in the future.

VI. REFERENCES

[1] N. Wirth, "A Brief History of Software Engineering", IEEE Annals of the History of Computing, 2008.
[2] M. Jackson,"Principles of Program Design", Academic Press, London, 1975.
[3] O. J. Dahl, C. Hoare, "Hierarchical program structures", Structured Programming, Academic Press, 1972, pp175-220.
[4] A. Snyder, "The essence of objects: Common concepts and termi-nology", Technical Report HPL-91-50, Hewlett Packard Laboratories, 1991.
[5] S. Alahmari, E. Zaluska, D. De Roure, "A Service Identification Frame- work for Legacy System Migration into SOA", 2010 IEEE

International Conference on Services Computing, 2010, pp614-617.

[6] N. Dragoni, S. Giallorenzo, A. L. Lafuente, M. Mazzara, F. Montesi, R. Mustan, L. Sana, "Microservices: yesterday, today, and tomorrow", https://arxiv.org/abs/1606.04036

[7] N. Dragoni, S.Dustdary, S.Larsenz, M.Mazzara, "Microservices: Migration of a Mission Critical System", https://arxiv.org/abs/1704.04173, 2017.

[8] E. Evans, "Domain-driven design: tackling complexity in the heart of software", Addison-Wesley Professional, 2004.

[9] M. Beer, W. Huang, R. Hill, "Designing community care systems with AUML". In: Proceedings of the International Conference on Computer, Communication and Control Technologies (CCCT '03), IEEE Computer Society, 2003, 247-253.

[10] R. Hill, S. Polovina, M. Beer, "From concepts to agents: towards a framework for multi-agent system modelling". In Proceedings of the fourth international joint conference on Autonomous agents and multiagent systems (AAMAS '05). ACM, New York, NY, USA, 2005, 1155-1156. DOI=http://dx.doi.org/10.1145/1082473.1082670

[11] A. Ikram, A. Anjum, R. Hill, N. Antonopoulos, L. Liu, S. Sotiriadis, "Approaching the Internet of things (IoT): a modelling, analysis and abstraction framework". Concurrency and Computation: Practice and Experience, 27(8), 2015, pp1966-1984.

[12] N. Bessis, F. Xhafa, D. Varvarigou, R. Hill, M. Li, "Internet of Things and Inter-cooperative Computational Technologies for Collective Intelligence". Studies in Computational Intelligence 460, Springer, 2013, ISBN 978-3-642-34951-5.

[13] R. Hill, J. Devitt, A. Anjum, M. Ali, "Towards In-Transit Analytics for Industry 4.0". FCST2017, IEEE Computer Society, 2017, Exeter.

[14] S. Newman, "Building Microservices - Designing Fine-Grained Systems", O'Reilly, 2017.

[15] D. Bryant, "The Seven Deadly Sins of Microservices". https://opencredo.com/the-seven-deadly-sins-of-microservices-redux/, 2016.

[16] J. Thones, "Microservices". IEEE Software, 32(1), 2015, pp116-116.

[17] N. Viennot, M. Lecuyer, J. Bell, R. Geambasu, J. Nieh, "Synapse: A Microservices Architecture for Heterogeneous-Database Web Applications". Proceedings of the Tenth European Conference on Computer Systems, ACM 2015.

[18] L. Walker, "IBM business transformation enabled by service-oriented architecture". IBM Systems Journal, 2007, 46(4).

[19] J. Lewis, M. Fowler, "Microservices a definition of this new architectural term". https://martinfowler.com/articles/microservices.html, 2014.

[20] E. Evans, "DDD & Microservices: At Last, Some Boundaries!". GOTO 2015.

[21] E. Evans E., "Domain-Driven Design, Tackling Complexity in the Heart of Software", Addison-Wesley, 2003, ISBN 0-321-12521-5.

[22] W. Hasselbring, "Microservices for Scalability: Keynote Talk Abstract". Proceedings of the 7th ACM/SPEC on International Conference on Performance Engineering Pages 133-134. 2016.

[23] 0MQ: Distributed messaging. http://zeromq.org/

[24] H.A.Aqrabi, L.Liu, R.Hill, N.Antonopoulos, "Taking the Business Intelligence to the Clouds". 9th International Conference on Embedded Software and Systems (HPCC-ICESS), Liverpool, IEEE Computer Society, 2012, pp953-958.

[25] Berners-Lee T. "Web Services" https://www.w3.org/DesignIssues/WebServices.html

[26] G. Alonso, F. Casati, H. Kuno, and V. Machiraju, "Web Services: Concepts, Architectures, Applications", Springer, 2004.

[27] S. Weerawarana, F. Curbera, F. Leymann, T. Storey, D. Ferguson, "Web Services Platform Architecture", 2005, Prentice Hall.

[28] X. Xu, L. Zhu, Y. Liu, M. Staples, "Resource-Oriented Business Process Modeling for Ultra-Large-Scale Systems". Proceedings of the 2nd international workshop on Ultra-large-scale software-intensive systems, 2008, pp65-68.

[29] Inc. Pivotal Software. "Rabbitmq - messaging that just works". https://www.rabbitmq.com/

[30] Inc. Kubernetes, "Kubernetes - production-grade container orchestration". http://kubernetes.io.

[31] Inc. Docker, "Swarm mode overview - docker". https://docs.docker.com/engine/swarm/.

[32] M. C. MacKenzie, K. Laskey, F. McCabe, P. F. Brown, R. Metz, B. A. Hamilton, "Reference model for service oriented architecture 1.0". OASIS Standard, 2006, 12.

[33] Inc. Mesosphere. "Marathon: A container orchestration platform for mesos and dc/os". https://mesosphere.github.io/marathon/.

[34] A. Mos, T. Jacquin, "A Platform-Independent Mechanism for Deployment of Business Processes Using Abstract Services", 17th IEEE International Enterprise Distributed Object Computing Conference Workshops, 2013, pp72-78.